\documentstyle{mn}

\newcommand{\lta}{{\small\raisebox{-0.6ex}{$\,\stackrel
{\raisebox{-.2ex}{$\textstyle <$}}{\sim}\,$}}}
\newcommand{\gta}{{\small\raisebox{-0.6ex}{$\,\stackrel
{\raisebox{-.2ex}{$\textstyle >$}}{\sim}\,$}}}  
\newcommand{\ale}{\ \raisebox{-.3ex}{$\stackrel{<}{\scriptstyle \sim}$}\ }

\newcommand{\msol}{{\rm M_\odot}}

\input psfig
\title[Superoutbursts In Dwarf Novae]
      {On The Nature Of Superoutbursts In Dwarf Novae}
\author[M. R. Truss et al]
{M. R. Truss, J. R. Murray and G. A. Wynn\\  
Department of Physics and Astronomy, University of Leicester,
University Road, Leicester, LE1~7RH, UK}
 	
\begin{document}
\maketitle
\begin{abstract}
We present the first detailed hydrodynamic simulation of a
superoutburst to incorporate the full tidal potential of a binary system.
A two-dimensional smoothed particle hydrodynamics code is used to
simulate a superoutburst in a binary with the parameters of the
SU UMa system Z Chamaeleontis. The simulated light curves shows all the
features observed in such systems.
Analysis of the mass flux through the disc and the growth rate of the
superhumps and disc eccentricity show that the superoutburst-superhump phenomenon is a direct 
result of tidal instability. No enhanced mass transfer from the
secondary is required to initiate or sustain these phenomena.
Comparisons of superoutbursts with normal outbursts are made and we
show that the model can be reconciled with the behavior of U Geminorum
type dwarf novae, which show no superoutbursts.
\end{abstract}
\section{Introduction}
SU Ursae Majoris systems are a class of cataclysmic variable which lie
below the period gap (they have $P_{{\rm orb}}<2.1$ hours) and which differ
in their outburst behaviour from dwarf novae in two distinct ways : in
addition to the normal outbursts displayed by dwarf novae which
last two to three days, SU UMa systems undergo longer
superoutbursts. These show a plateau of brightness extending the
outburst to ten days or more. Also, during a superoutburst, the light
curve of an SU UMa reveals superhumps - a superimposed variation in
brightness with a period a few percent longer than the orbital period of the
system. It has been suggested that normal and superoutbursts do not coexist
independently of each other, with a superoutburst being initiated by a
normal outburst. There is evidence for a relaxation time of $\sim 170$ days
in VW Hyi (van der Woerd and van Paradijs 1987), with all normal
outbursts triggering a superoutburst if they occur 170 days or longer
after the previous superoutburst. 
\subsection{Outburst mechanisms}
Historically, two models have been proposed to explain dwarf nova
outbursts : a varying mass transfer rate from the secondary due to  
its envelope being dynamically unstable (Bath 1973), or a varying
viscosity in the accretion disc between quiescent and outburst states
(Hoshi 1979). It is the disc instability model that has prevailed as
the accepted explanation, offering better compatability with modern
observations and theory.  An excellent review is given by Cannizzo
(1993). We have recently used a smoothed particle hydrodynamics (SPH)
model to simulate a wide range of dwarf nova outburst characteristics
(Truss et al., 2000).\\\indent
A complete theory of superoutbursts has been much more elusive.
Several models have been presented, however few explain all the 
necessary properties, in particular the superhumps which appear
without fail in every superoutburst. Models involving mass-transfer
instabilities (Osaki 1985), disc instabilities (Osaki 1989) and both
(Duschl and Livio 1989) all explain some properties of a superoutburst.
A brief summary of these can be found in Warner (1995).
Simulations by Whitehurst (1988) showed that discs in binaries with
mass ratios $q \ale 1/4-1/3$ could become tidally unstable. Unlike
stable discs which had a shape that was fixed in the binary frame,
these unstable discs were significantly eccentric. In the inertial
frame this eccentricity executed a slow prograde precession. Thus in
the binary frame the disc precessed retrogradely. Whitehurst found
that the tidal stresses on the precessing disc gave rise to a
periodically varying light curve that neatly explained superhumps. 
The disc instability was explained in detail by Lubow (1991a) in terms of an
(eccentric inner) Lindblad resonance that occurs where orbits in the
disc are $3:1$ resonant with the binary orbit.
Only discs in systems with mass
ratios $q < 1/4- 1/3$ are large enough to reach the resonance.
This theoretical result closely matches the spread of cataclysmic
variables that are observed to have superhumps (Patterson, 1998).\\\indent
Osaki (1989) espoused a superoutburst model which combined the thermal
and tidal instabilities. His model relied on the the outer disc radius
varying significantly over the  course of a superoutburst cycle. At
the end of  a superoutburst, Osaki proposed that the disc radius was
small so that it could not access the $3:1$ resonance. Hence when the
disc became thermally unstable, a normal dwarf nova outburst
occurred. As a result of the outburst however, the disc spread
radially.  With each successive normal outburst the disc grew until
it encountered  the $3:1$ resonance and became tidally as well as
thermally unstable. Osaki proposed but could not conclusively show,
that the tidal removal of angular momentum was much more efficient
from an eccentric disc. Consequently, when the disc encountered the
resonance it would dump a large fraction of its mass upon the white
dwarf. A prolonged ``super'' outburst that left a much diminished
disc resulted. Osaki's model could not be carried too far as the
efficiency with which tides removed angular momentum from the outer
disc could not be determined with certainty. Osaki, and other
workers, were relying upon one dimensional numerical models for disc
evolution in which the tidal forces had only been included
approximately.\\\indent 
In this paper we report the results of a smoothed particle
hydrodynamics (SPH) simulation of a cataclysmic variable binary with
mass ratio $q=M_{{\rm 2}}/M_{{\rm 1}}=0.15$, representative of
the SU~UMa system Z Chamaeleontis. As the code is fully three
dimensional (though for the purposes of this paper the third dimension
has been suppressed), tidal forces are included exactly. Thus for the
first time we are able to follow the evolution of an accretion disc over the
course of a superoutburst.
\section{Numerical Method}
\subsection{The Model}
The model and methods used in this work have already been
described in our recent paper (Truss et al., 2000), as applied to a
system with the properties of SS Cygni (0.6).
The SPH code directly includes the tidal forces due
to the secondary, and incorporates a simple model for the dwarf nova
thermal instability.

The key to our calculations lies in the implementation of a
viscosity switch. If the local surface density exceeds a defined value
$\Sigma_{\rm max}$
then that region is transformed to the high-viscosity outburst state
on a time-scale appropriate to the thermal time-scale. There is a
corresponding second trigger level, $\Sigma_{\rm min}$, which causes
the viscosity to be switched back down to the quiescent level. In this
way we can simulate the limit-cycle behaviour of a dwarf nova with an
isothermal simulation. Viscosity switches are nothing new in themselves in disc
simulations, but in the past they have used trigger levels which are constant
throughout the disc. This is an unrealistic situation. Cannizzo,
Shafter and Wheeler (1988) have calculated the format for the critcal
values of surface density $\Sigma$ in a steady-state disc and found an
almost linear relationship with disc radius R (Cannizzo, Shafter and
Wheeler, 1988, equations 2.1 and 2.2). Therefore, we implement a
linear $\Sigma - R$ relationship for the disc in steady-state.

There are, however, some important refinements that have been made for
this study which were not made previously. In Truss et al. (2000) we
incorporated an azimuthally smoothed trigger condition, in which the
density condition was only tested for annuli in the disc. This is
unsatisfactory for work on a system with a more extreme mass ratio
because the outer regions of the disc depart strongly from azimuthal
symmetry in response to the tidal resonance. We therefore calculate
our trigger condition absolutely locally for every particle in the
simulation. We also previously used an enhanced viscosity parameter
($\alpha = 0.1$ in quiescence and $\alpha = 1.0$ in outburst) to
improve the run-time of the code. The desire to examine the
growth-rate of superhumps on the correct time-scale has motivated us
to use more realistic viscosity parameters here at the expense of
simulating many outburst cycles. We use $\alpha = 0.01$ in quiescence
and $\alpha = 0.1$ in outburst, typical of the values in a dwarf nova
disc (see, Shakura \& Sunyaev (1973) for a discussion of this
viscosity parameterisation). We use other physical parameters relevant
to observations of Z Cha - $\rm P_{orb} = 0.075~d, \dot{M}_2 = 2 x
10^{15} gs^{-1}$ (both from Wood et al. 1986) and $\rm M_1 = 0.84
\msol$ (Wade et al. 1988). The disc is built up from a mass stream
and undergoes normal outbursts during the build-up to steady state in
exactly the same way as our previous simulations (Truss et
al. 2000).
\section{Results}
\subsection{Light Curves} 
We present light curves through the entire superoutburst. The curves are
constructed by summing the viscous dissipation in different regions of
the disc. In this way we are able to compare the results with observed 
lightcurves in different wavebands. This approach differs slightly
from our previous work, in which we attempted to reconstruct several
wavebands by treating the disc as a black body. Since the observed optical
emission is expected to be dominated by the cooler outer regions of the disc and the
extreme ultraviolet (EUV)
emission is dominated by the very hot inner region, we can gain a 
qualitative understanding of the behaviour of the disc with this
simplified approach. A more quantitative analysis should be left to future
work in which full thermodynamics is incorporated
self-consistently. 
\begin{figure}
\psfig{file=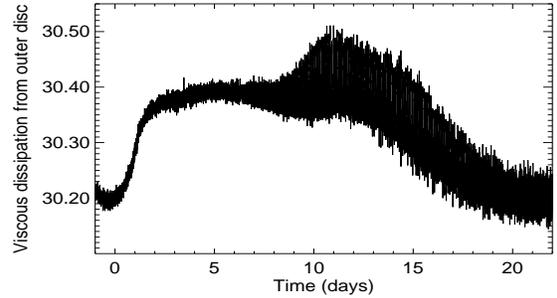,width=8cm,height=4.5cm}
\caption{Viscous dissipation from disc radii $\rm R > 20R_{wd}$. This
is representative of the observed visual lightcurve. The scale of the
dissipation program units is logarithmic and t=0 represents the
beginning of the outburst.} 
\label{ubv}
\end{figure}  
Figure~\ref{ubv} shows the viscous dissipation from the outer parts of
the disc (R \gta 20 $R_{\rm wd}$). The profile and duration of the outburst is in
good agreement with V-band observations of SU UMa systems. The initial rise
to outburst is fairly rapid, lasting for two days, and is followed by
the characteristic superoutburst plateau. Superhumps begin to appear
prominently at t = 8 days.  The decay of the superhumps in the
simulation is slower than observations have suggested - they persist
(albeit with decreasing amplitude)  during and just after the decline from
supermaximum. Buat-M\'{e}nard et al. (2000) included the effects of
stream-impact heating in a one-dimensional model, and showed that this
causes the functional form of the critical surface density to rapidly
decrease at the disc edge. We believe that this would cause the outer
disc to drain faster than in our simulation and drive the edge away from
the resonance more quickly, suppressing the superhumps sooner. 
The decay from supermaximum is slower than the rise, lasting
for 5 days. The total duration of the simulated superoutburst is
around 18 days.
\begin{figure}
\psfig{file=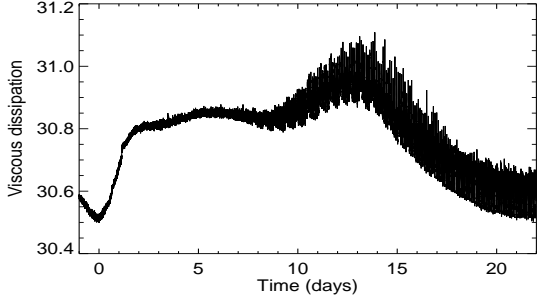,width=8cm,height=4.5cm}
\caption{Viscous dissipation from $\rm R < 10R_{wd}$. There is a
noticeable rebrightening around t = 10 days. }
\label{diss}
\end{figure}
\begin{figure}
\psfig{file=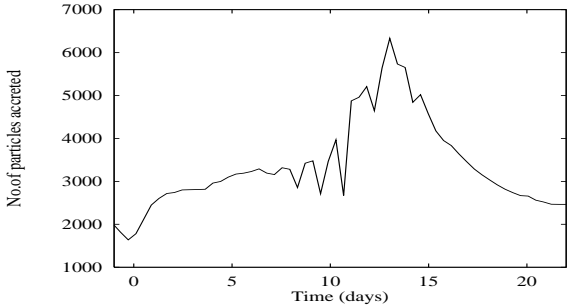,width=8cm,height=4cm}
\caption{Mass accretion rate onto the primary.}
\label{xray}
\end{figure}
Recent EUVE observations of OY Carinae, which has a similar mass ratio
to Z Cha, have shown that in  addition to the expected delay between
the optical and EUV rise, the EUV emission shows a rebrightening
during superoutburst which is not seen in the V-band (Mauche et al.,
2000). We are able to look for this effect in the simulation by
consideration of the viscous dissipation coming from the hot, inner
part of the disc (R \lta 10 $R_{\rm wd}$), which is shown in
Fig.~\ref{diss} and the accretion rate onto the white dwarf (a good
measure of the X-Ray lightcurve - Fig.~\ref{xray}). The rebrightening
can clearly be seen in both figures, and is caused by the arrival of
additional material from the outer disc which has been tidally forced
into the high state.  
\subsection{Disc Response}
It is particularly instructive to construct a visualisation of the
disc at various times through the superoutburst cycle. 
\begin{figure*}
~~\psfig{file=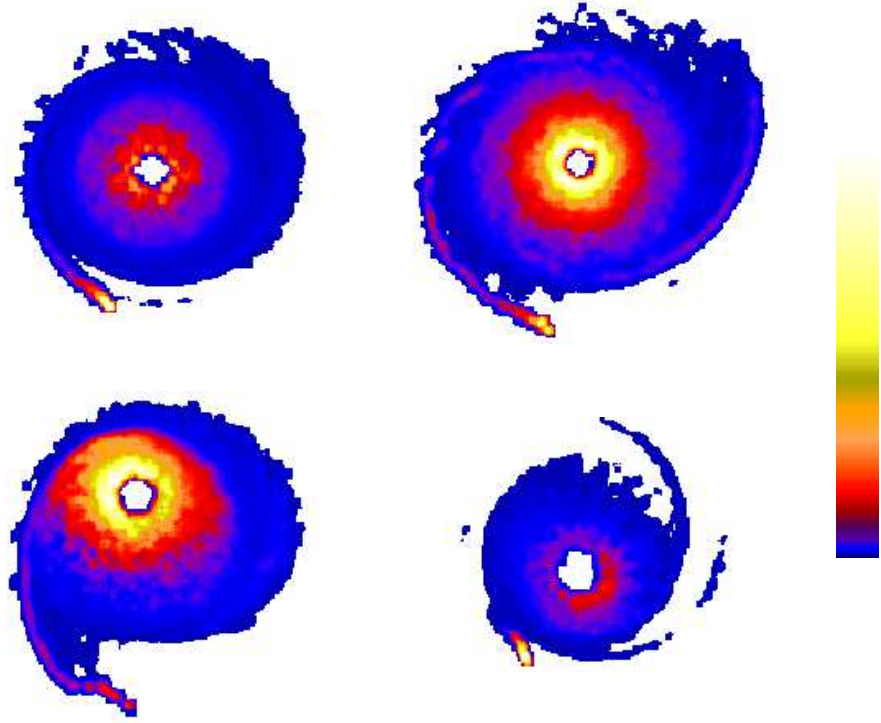,width=12cm}
\caption{Viscous dissipation from the disc. Top left : quiescence
(t=0 days); top right : rise to outburst before contact with the
resonance (t=3 days); bottom left : superoutburst (t=13 days); bottom
right : after superoutburst (t=23 days). The colour scaling is
logarithmic from $10^7\rm{ergs^{-1}cm^{-2}}$ to
$10^9\rm{ergs^{-1}cm^{-2}}$. The white central portion of the disc ($R
< 4 R_{\rm wd}$) is not modelled in the simulation.} 
\label{discs}
\end{figure*}
Figure~\ref{discs} shows dissipation maps of the disc in four
states. The hot-spot and spiral structure of the disc are clearly
visible. In quiescence, the disc is not in contact with the 3:1
resonance; consequently it is circular. The innermost part of the disc
remains in the high state. This is in agreement with our previous
work on SS Cygni. The outburst progresses just as for a normal
outburst, but during this time the disc radius increases and the edge
of the disc encounters the resonance. We stress that this resonance is 
not available to systems with less extreme mass ratios. The disc
rapidly becomes eccentric and virtually the entire disc is transformed 
to the high viscosity outburst state. It is interesting to note the
brightness of the hot-spot in the lower-left hand panel of
fig.~\ref{discs}. During supermaximum, we find that the brightness of
the hotspot varies according to the orientation of the eccentric disc. 
This is not due to a varying mass transfer rate from the secondary -
this remains constant at all times during the simulation - it is
purely a result of the geometry of the stream impact region. The
existence of such an effect in the nova-like V348 Pup has  been
reported by Rolfe et al. (2000). 
\begin{figure}
\psfig{file=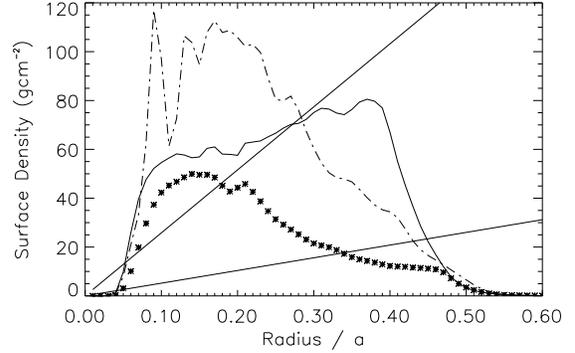,width=8cm}
\caption{Surface density evolution of the disc.The plots correspond to
times t=5.0 days (solid line),t=10.3 days (dot-dashed line) and t=18.1
days (stars). The straight lines are the critical surface density
conditions $\Sigma_{\rm max}$ and $\Sigma_{\rm min}$ from the S-curve
disc instability model. a is the binary separation}
\label{dens}
\end{figure}
Figure~\ref{dens} shows the evolution of surface density with disc
radius. Initially, the inner regions (R \lta 0.25a) go into outburst (solid
line), but it should be noted that the large reservoir of mass at
larger radii is very close to the upper critical
threshold. Consequently, it is only a short time before nearly the
whole disc is in outburst, and at supermaximum the large reservoir of 
mass is seen to move to smaller radii (diamonds). At the end of
the superoutburst, a large fraction of the disc has been accreted
(stars); $\sim 60 \%$, compared with $\sim 5 - 10 \%$ for a normal 
outburst.  
\subsection{Modal Analysis and Superhumps}
We compare the response of the disc to the tidal field of the
secondary star with the analysis of Lubow (1991a). The tidal potential can
be decomposed into a set of functions 
\begin{equation}
\phi(r,\theta,t) = \sum_{m=0}^{\infty}
\phi_m(r)cos[m(\theta-\Omega_{\rm orb}t)],
\end{equation}
\noindent
where each mode m generates a sinusoidal response in the disc with
argument ($k\theta-l\Omega_{\rm orb}t$). The (k,l) = (1,0) mode represents the
disc eccentricity, which grows exponentially at a rate
\begin{equation}
\lambda = 2.08 \times 2\pi\Omega_{\rm orb}q^2r^2_{\rm res}{\Sigma(r_{\rm
res})\over M_{\rm disc}}.
\end{equation}
\noindent
$\Sigma(\rm r)$ is the surface density of the disc at radius r,
$M_{\rm disc}$ is the mass of the disc and $r_{\rm res}$ is the radius
of the Lindblad resonance. This resonance couples with any existing
eccentricity to launch a two-armed travelling spiral wave with (k,l) = (2,3). 
\begin{figure}
\psfig{file=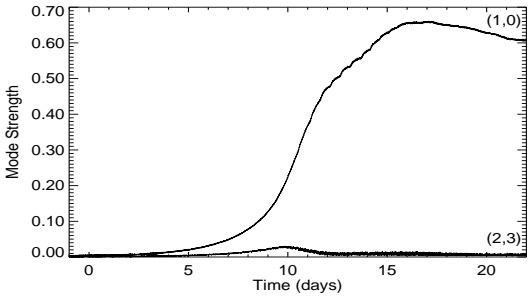,width=8cm,height=4.5cm}
\caption{Relative mode strengths. The edge of the disc encounters the
3:1 resonance around t = 5 days, producing a rapid growth in the
eccentric mode (1,0). The (2,3) mode represents a travelling spiral
wave launched at the resonance.} 
\label{modes}
\end{figure}
The response of these two modes in our simulation is shown in
Fig.~\ref{modes}. It is immediately clear that, as Lubow predicted, 
the growth of eccentricity is proportional to the strength of the
(2,3) mode. For times less than t=11 days in the simulation, the
growth rate of the (1,0) mode was
$$
\lambda = 0.0075 \pm 0.0002 \Omega_{\rm orb}
$$
\noindent
which compares favourably with the analytical prediction for this system of 
$$
\lambda \simeq 0.01 \Omega_{\rm orb}.
$$ 
\noindent 
\begin{figure}
\psfig{file=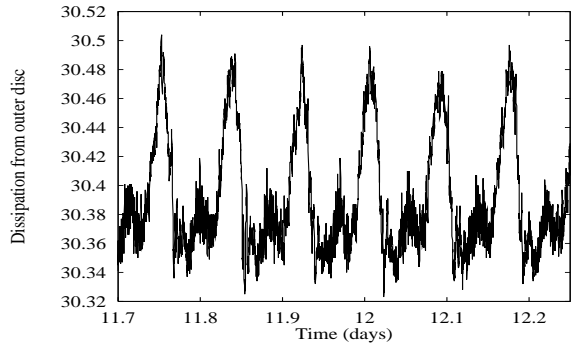,width=8cm,height=4.5cm}
\caption{Superhumps in lightcurves from the outer disc.}
\label{sh}
\end{figure}
Let us turn now to the superhumps themselves. The superhump period is,
as expected from observations, slightly longer than the
orbital period, and gradually decreasing as the outburst progresses. A
Fast Fourier Transform of the dissipation data in the range where the
eccentricity growth is a maximum (9.5 - 12 days) yields the period
$$
P_{\rm sh} = 1.039 \pm 0.008 P_{\rm orb}.
$$
\noindent
However, between 12 and 18.5 days, when the superhumps are fully
developed, we obtain  
$$
P_{\rm sh} = 1.029 \pm 0.003 P_{\rm orb}.
$$
\noindent
These results are in excellent agreement with the observations
collated by Warner \& O'Dononghue (1988), who found the superhump period to
decrease according to 
$$
P_{\rm sh} = (1.0389 - 0.0007T) P_{\rm orb}
$$
\noindent
where T is the time after the beginning of the outburst in days.
\begin{figure}
~~~~~~~~~~~~~~~\psfig{file=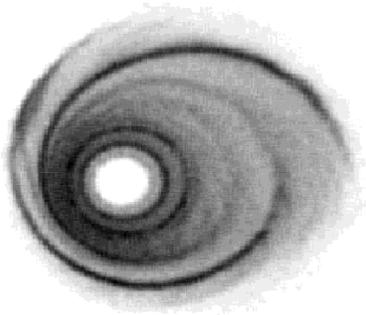,width=5cm}
\caption{Surface density plot of the disc showing the spiral wave launched on contact with the resonance.}
\label{shdisc}
\end{figure}
\noindent
Fig.~\ref{sh} shows a 15 hour portion of the lightcurve from the same
region of the disc as defined in fig.~\ref{ubv}. We also find a
strong superhump signal produced from the inner part of the disc. This
emission results  from the spiral shock wave reaching all the way down
into this region, and the effect can be seen with striking clarity in
the map of the disc in fig.~\ref{shdisc}.  
\section{Discussion}
We have performed the first two-dimensional simulation of a dwarf nova
superoutburst. The method includes the full tidal potential of the
binary and reproduces the observed characteristics of a
superoutburst with no need for enhanced mass transfer from the
secondary star. This leads us to conclude that the
superoutburst-superhump phenomenon is purely a result of the tidal
instability working in tandem with the thermal disc
instability. Superoutbursts are therefore not seen in systems with less
extreme mass ratios ($q \gta 0.25$) because the tidal resonance
is not available to them.

We feel that there is much confusion in the literature regarding the
terminology of 'wide' and 'narrow', 'inside-out' and 'outside-in'
outbursts. This work coupled with our previous study of normal
outbursts suggests the following :\\
1. An outburst may be initiated at any point in the disc and in
   general the resultant heating wave will travel in {\em both directions}.\\
2. All types of dwarf nova can exhibit both wide and narrow outbursts.
   The physical distinction between the two is simply whether the
   heating wave propagates to the large reservoir of mass stored in
   the outer parts of the disc or not.\\
3. The superoutburst/superhump phenomenon is a result of the presence
   of the tidal instability, {\em in addition} to the thermal
   instability. The tidal instability is responsible for superhumps,
   and the duration of a superoutburst is primarily determined by the
   hot region of the disc arising from the thermal
   instability. However, the duration will be lengthened by the
   enhanced mass accretion arising from the tidally heated gas, which
   produces the EUV rebrightening discussed in section 3.1.\\
\noindent
Long-term observations of dwarf novae are still needed across a
variety of wavebands in order to improve our understanding of these
objects. EUV and X-Ray observations will reveal much about the physics
of the boundary layer between the disc and the surface of the white
dwarf, while eclipse mapping and d\"{o}ppler tomography should probe the
structure of the disc and the r\^{o}le of the spiral waves.  
We also point out that observations of the growth rate of superhumps can
be used to gain estimates for mass ratio, through referral to
Lubow's modal analysis that we have reprised here.

It is very important to concentrate on two and three dimensional
simulations in the future, as the full tidal field is inherent in
these models. The next step should be to perform a 3D calculation
which includes fully self-consistent thermodynamics. One can envisage
a hybrid of a 2D isothermal code such as this one with a traditional 1D
thermodynamics code. This is a mammoth task in terms of computational
requirements, but all the components of such a model already exist.  
\section*{Acknowledgments}
Research in theoretical astrophysics at Leicester is supported by a
PPARC rolling grant. The simulations were performed using GRAND, a
high-performance computing facility based at Leicester and funded by
PPARC. MRT acknowledges a PPARC studentship and the support of the
William Edwards Educational Charity.  

\end{document}